\documentstyle[aps,multicol,epsf]{revtex}

\def\twobeone{\end{multicols}
\vskip.6pc
\noindent
\vrule width3.375in height.2pt depth.2pt \vrule depth0em height1em\hfill 
\vskip.6pc
\widetext}
\def\onebetwo{
\vskip.6pc
\indent
 \hfill\vrule depth1em height0pt \vrule width3.375in height.2pt depth.2pt
\vskip.6pc
\begin{multicols}{2}\narrowtext \noindent}
\begin{document}
\draft
\title{Crossover Phenomena in the One-Dimensional
SU(4) Spin-Orbit Model \\under Magnetic Fields}

\author{Yasufumi Yamashita, Naokazu Shibata\cite{address}, 
and Kazuo Ueda}
\address{
Institute for Solid State Physics, University of Tokyo, 
Roppongi 7-22-1, Minatoku, Tokyo 106-8666, Japan
}
\date{Received \today}
\maketitle
\begin{abstract}
We study the one-dimensional SU(4) exchange model under magnetic fields,
which is the simplest effective Hamiltonian in order to investigate 
the quantum fluctuations concerned with the orbital degrees of freedom 
in coupled spin-orbit systems.
The Bethe ansatz approaches and numerical calculations using 
the density matrix renormalization group method are employed.
The main concern of the paper is how the system changes from the SU(4) 
to the SU(2) symmetric limit as the magnetic field is increased.
For this model the conformal field theory predicts an usual behavior: 
there is a jump of the critical exponents just before the SU(2) limit.
For a finite-size system, however, 
the orbital-orbital correlation functions 
approach continuously to the SU(2) limit after interesting
crossover phenomena.
The crossover takes place in the magnetization range of
1/3 $\sim$ 1/2 for the system with 72 sites studied in 
this paper.
\end{abstract}
\begin{multicols}{2}
\narrowtext

\section{Introduction}\label{secsu4_field}
In recent years increasing attention has been paid to the physics
related to the orbital degrees of freedom and its quantum fluctuations 
in the strongly correlated electron systems
in addition to the charge and/or spin degrees of freedom.
In the conventional treatments for the coupled spin and orbital systems,
mean-field-type approaches have been used to determine an 
ordered orbital structure and the effects of the quantum fluctuations
are usually neglected\cite{kk,inagaki}.
Recent experiments suggest, however,
in some transition metal and rare earth compounds,
the orbital degeneracies and quantum fluctuations 
of spin and orbital degrees of freedom 
play an important role in their unusual properties,
which may require more sophisticated treatments than the
mean-field approximations.

For examples, CeB$_6$ is known as a typical dense Kondo system 
($T_K=1\sim 2$ K) with the $\Gamma _8$ quartet ground state, 
and is well known for its unique
magnetic field-temperature phase diagram\cite{ceb}. 
Since the quadrupolar ordering
temperature ($T_Q\sim 3.4$ K) 
is of the same order as the spin N\'{e}el 
ordering temperature ($T_N\sim 2.3$ K),
the interplay between spin and orbital quantum fluctuations 
may be important and thus we need to consider
the effects of quantum fluctuations more seriously
beyond the mean-field theories.
A pure LiV$_2$O$_4$,
which has a fcc normal-spinel structure and
does not show any static magnetic order,
is found to be the first compound which 
shows the heavy fermion 
like behavior among the transition metal oxides\cite{liv2o4}.
But its origin of the large effective mass 
may be different from the usual heavy-fermion
systems which have nearly localized $f$- and itinerant conduction 
electrons. 
The highly frustrated lattice structure and
nearly degenerate $t_{2g}$ orbital at each V$^{3.5+}$ ion
are believed to be the origin of the unusual properties, such as
disordered ground state and heavy-fermion like behaviors.

Generally speaking, the quantum fluctuations are less important
in higher dimensions, and the classical pictures, i.e. spin and
orbital ordered structures in the case of the couples spin and orbital
models, may be applied for many cases in three dimensions.
However, these experiments suggest the possibilities that
the interplay between spin and orbital fluctuations destroys 
the classical ordered structures and 
the ground state may become a disordered liquid state.

In order to study the physics in which 
quantum fluctuations concerned with the orbital degrees 
of freedom are important, 
the SU(4) exchange Hamiltonian (\ref{su4exchange})
has been considered as a prototypical model
and studied extensively by several groups\cite{yamashita,FuChun,troyer}. 
This SU(4) exchange model is derived from the 
highest symmetric limit of the 
two band Hubbard model 
with the strong correlation ($U/t \gg 1$) and 
the vanishing Hund coupling ($J=0$) at quarter filling.
The model is given by,
\begin{eqnarray}
 H_{\rm{SU(4)}}=K\sum_{\langle i,j \rangle}
\left(2\vec{S}_{i}\vec{S}_{j}+\frac{1}{2}\right)
\left(2\vec{T}_{i}\vec{T}_{j}+\frac{1}{2}\right),\label{su4exchange}
\end{eqnarray}
where $\langle i,j\rangle$ stands for a nearest neighbor
pair and $K$ is given by $2t^2/U>0$.
It consists of usual $\vec{S}$- and pseudo $\vec{T}$- spins, 
which describe the spin and orbital degrees of freedom, respectively.
It is important to note that the both spins have the Heisenberg 
type isotropic interactions
and the total symmetry of this model is the SU(4) group, which is 
higher than the apparent SU(2)$\otimes$SU(2) group.

In one dimension,
overall properties of the ground state and excitations 
of this most symmetric SU(4) exchange model
are well understood by the previous studies
\cite{yamashita,troyer,FuChun2,bethe,affleck}.
In terms of the spin-ladder models, Eq.(\ref{su4exchange})
corresponds to the two-leg AF Heisenberg ladder with 4-body
interaction producing frustrations.
A similar model without frustration (the 4-body
interaction with the opposite sign) is 
studied by using the green's function Monte Carlo method\cite{negativek}.
It is noteworthy that the present
model belongs to the Bethe solvable class with the general
SU($N$) symmetry and also
belongs to the quantum critical systems 
with strong quantum fluctuations.

In actual systems, the highest SU(4) symmetry
may be destroyed into various lower symmetries 
due to some anisotropies and 
different physics emerges depending on the way 
how the symmetry is lowered.
In many physically relevant systems, 
the spin rotational invariance is approximately hold,
while the SU(2) symmetry of the orbital part
is easily destroyed. For example, 
it becomes U(1) by a single-ion anisotropy 
and a finite Hund coupling makes 
the exchange interaction of the orbital part Ising type
with Z$_2$ symmetry\cite{yamashita}.
Another type of symmetry lowering is the models 
with SU(2)$\otimes$SU(2) 
symmetry which is given by Eq.(\ref{su4exchange})
with arbitrary constants replaced for 1/2.
This model in one dimension is investigated both numerically and 
analytically by several authors\cite{singh,nersesyan}.

In this paper we study the one-dimensional SU(4) exchange model 
under magnetic fields as another example with a lower symmetry.
It is naturally expected that the quantum fluctuations in one
dimension are more prominent than those in two or three
dimension. Therefore one dimensional model may shed right on
the role of quantum fluctuations concerned with the
spin and orbital degrees of freedom.
It should be also mentioned that this model is equivalent 
to the case with a splitting between the two orbitals.
Under magnetic fields,
the highest SU(4) symmetry is broken by the Zeeman splitting and
the total symmetry of the model becomes SU(2)$\otimes$U(1), 
where the SU(2) symmetry represents the rotational 
invariance of the orbital degrees of freedom 
and the U(1) symmetry describes the spin degrees of freedom
under uniaxial magnetic fields.
In sufficiently strong magnetic fields,
it is evident that the all spins are 
fully polarized and our model reduces to the 
SU(2) symmetric antiferromagnetic (AF) Heisenberg model with only
pseudo $\vec{T}$- spin degrees of freedom.

Therefore,
it is interesting to investigate how the system changes 
from the SU(4) to the SU(2) symmetric limit 
as we increase the magnetic field. 
We study this transition by calculating 
the ground-state energy and elementary excitations and 
also by looking at the behaviors of 
the orbital-orbital correlation functions.
For this purpose,
we have used numerical calculations with the density matrix
renormalizing group (DMRG) method as well as analytic 
approaches based on the Bethe ansatz and 
the conformal field theory (CFT).  
Since these methods do not introduce any uncontrolled approximations,
they are suitable for investigating the effects of quantum fluctuations 
beyond the level of various mean-field theories. 

\section{Ground state and excitations}
\subsection{Bethe ansatz equations and symmetry properties}
We consider the one dimensional SU(4) exchange model 
with the periodic boundary conditions (PBC).
First we calculate the thermodynamic ground-state energy 
under magnetic fields by using
the Bethe ansatz methods for finite and infinite size systems.
In the actual calculations, we solve the Bethe ansatz
equations with specified values of the magnetization $m_z$, 
which is defined by $2S_{tot}^z/N$.
In this notation, $m_z =0$ corresponds to the
SU(4) symmetric limit with no magnetic field and 
$m_z=1$ corresponds to the SU(2) symmetric limit
where all $\vec{S}$- spins order in parallel and 
only orbital degrees of freedom are left.
The ground-state energies for the SU(4) and SU(2) symmetric limit
are obtained analytically to be 
$1-\pi/4-3\log{2}/2 \simeq -0.825\,118\,9$ and 
$1-2\log{2}\simeq-0.386\,294\,4$, respectively\cite{bethe}.

Let us suppose that $N_1$, $N_2$, $N_3$, and $N_4$ are the 
numbers of four species ($N=N_1+N_2+N_3+N_4$ and $N$ 
is the total number of sites) with the quantum numbers
$(S^z,T^z)=(\uparrow,\uparrow)$, $(\uparrow,\downarrow)$, 
$(\downarrow,\uparrow)$, and $(\downarrow,\downarrow)$, respectively. 
These numbers specify the value of $S_{tot}^z=(N_1+N_2-N_3-N_4)/2$ and 
$T_{tot}^z=(N_1-N_2+N_3-N_4)/2$.
We also define $M_1=N_2+N_3+N_4$, $M_2=N_3+N_4$, and $M_3=N_4$.
The Bethe ansatz equations for the SU(4) exchange model with 
the quantum numbers $N_1$, $N_2$, $N_3$, and $N_4$
are given as follows\cite{bethe}:
\twobeone
\begin{eqnarray}
NK(\alpha_i)&=&2\pi J_{\alpha_{i}}
-\sum_{i'=1}^{M_1}\theta \left(\alpha_i -\alpha_{i'} \right)
+\sum_{j=1}^{M_2}\theta \left(2\alpha_i -2\beta_j\right)\nonumber \\
0&=&2\pi J_{\beta_{j}}
-\sum_{j'=1}^{M_2}\theta \left(\beta_j -\beta_{j'} \right)
+\sum_{i=1}^{M_1}\theta \left(2\beta_j -2\alpha_i \right)
+\sum_{k=1}^{M_3}\theta \left(2\beta_j -2\gamma_k \right)\nonumber \\
0&=&2\pi J_{\gamma_{k}}
-\sum_{k'=1}^{M_3}\theta \left(\gamma_k -\gamma_{k'} \right)
+\sum_{j=1}^{M_2}\theta \left(2\gamma_k -2\beta_j\right),\label{finite_BA}
\end{eqnarray}
\onebetwo
where $K(\alpha)=2\tan^{-1}(2\alpha)\pm \pi$,
$\theta (x)=-2\tan^{-1}x$, and the suffix 
$i,j$, and $k$ are numbered from 1 to $M_1,M_2$ and $M_3$, respectively.
For a given set of $J_\alpha, J_\beta$, and $J_\gamma$ 
we can calculate the parameters $\alpha_i,\beta_j$ and $\gamma_k$,
so-called rapidities,
by solving the $M_1+M_2+M_3$ coupled non-linear equations (\ref{finite_BA}).
The quantum numbers $J_{\alpha}$, $J_{\beta}$ and $J_{\gamma}$
are integers (half integers) depending on the value
of the $M_1+M_2$, $M_1+M_2+M_3$, and $M_2+M_3$ 
to be odd (even), respectively, and their distributions 
specify a state.
The ground-state energy per site ($\varepsilon_N$),
momentum ($P$), and $m_z$ for the specified number of 
species, $N_1,N_2,N_3$, and $N_4$,
are given as follows:
\begin{eqnarray}
\varepsilon_N &=&1-\frac{2}{N}\sum_{i=1}^{M_1}
\Big{\{}1-\cos{K(\alpha_{i})}\Big{\}}=
1-\sum_{i=1}^{M_1}\frac{4}{1+4\alpha_i^2}\label{e},\\
P&=&\sum_{i=1}^{M_1}K(\alpha_{i})=\frac{2\pi}{N}\left(
\sum_{i=1}^{M_1}J_{\alpha_{i}}+\sum_{j=1}^{M_2}J_{\beta_{i}}+
\sum_{k=1}^{M_3}J_{\gamma_{i}} \right)\label{p} ,\\
m_z&=&n_1+n_2-n_3-n_4,\label{s}
\end{eqnarray}
where we take $K$ as energy units hereafter, putting $K=1$ in 
Eq.(\ref{su4exchange}), and define $n_i\equiv N_i/N$. 
In order to avoid complications coming from multi-fold 
degeneracies\cite{yamashita},
we only consider the system with $N=4n$ sites ($n$ is an integer) 
in the following.

\begin{figure}[htb] \begin{center}
\leavevmode
\epsfxsize=45mm
\epsffile{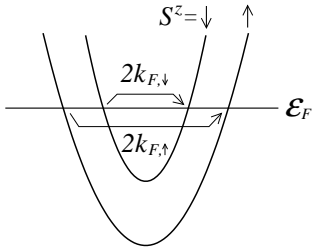}
\caption{ 
Schematic band structure of the SU(4) exchange model
under magnetic fields. $\varepsilon _{F}$ denotes the 
Fermi energy.}\label{band}
\end{center} \end{figure}

\begin{figure}[htb] \begin{center}
\leavevmode
 \epsfxsize=86mm
\epsffile{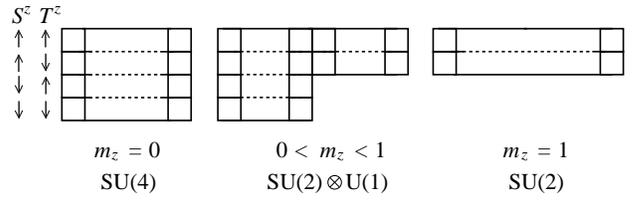}
\caption{
The Young's diagrams for the ground states under magnetic fields.
Below each diagram, 
magnetization and its total symmetry are given.
}\label{mag_gs}
\end{center}\end{figure}

Under magnetic fields, 
the four-fold degenerate bands in terms of the original Fermion model,
are split into two doubly degenerate bands by the Zeeman effect and 
these two bands are labeled by spin up and down, 
respectively (see Fig.{\ref{band}}).
Since $\vec{S}$- spins are gradually polarized 
with the increase of magnetic fields
and pseudo $\vec{T}$- spins are not affected directly,
it is reasonable to suppose that
the ground state in magnetic fields is represented by a
irreducible representation with a finite $S_{tot}^z$ and 
zero $T_{tot}^z$, which is given by the Young's diagram as
illustrated in Fig.\ref{mag_gs}.
For $m_z=0$ and 1, the irreducible representations of the ground state
are given by the SU(4) and SU(2) singlet, respectively.
For $0<m_z<1$, the upper and lower half rectangles represent the local
SU(2) symmetry of the orbital degrees of freedom for the up and down 
spin bands split under magnetic fields, respectively.

\subsection{Ground-state energies under magnetic fields}
\begin{figure}[htb]
\begin{center}
 \leavevmode
 \epsfxsize=76mm
\epsffile{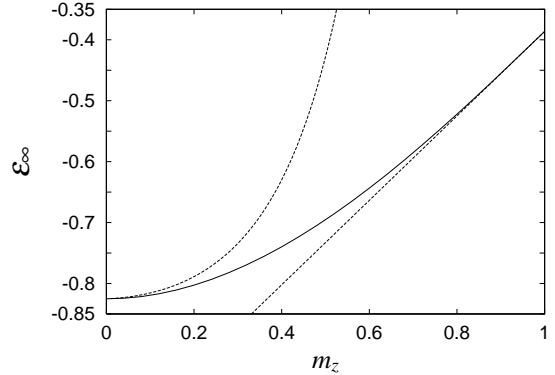}
\caption{
Thermodynamic ground-state energy as a function of $m_z$.
The solid line represents the ground state-energy obtained by the 
extrapolation to infinite system size.
Broken lines stand for the asymptotic behaviors 
around the SU(4) and SU(2) symmetric limits.
}\label{s_vs_e}
\end{center}
\end{figure}

We have numerically solved the coupled equations(\ref{finite_BA}) 
under the conditions, $N_1=N_2$ and $N_3=N_4$, and
by using Eq.(\ref{e}) and (\ref{s})
the ground-state energies 
with specified magnetization $m_z$ are calculated
for the system size $N=100$, $200$, and $400$, respectively.
In order to estimate the bulk limit, we have 
extrapolated to the infinite system size limit 
by fitting these data with the function 
$\varepsilon_N = \varepsilon_{\infty}+c_1/N+ c_2/N^2$. 
Bulk ground-state energies per site as a function of $m_z$ are obtained
as shown in Fig.{\ref{s_vs_e}} by the solid line.
The ground-state energy for the SU(4) and SU(2) 
symmetric limit are calculated 
to be $-0.825\,119\,0$ and $-0.386\,294\,4$, 
which are perfectly consistent with the analytic results.

In order to consider the thermodynamic properties of the SU(4) 
exchange model and to analytically obtain the asymptotic forms 
around the SU(4) and SU(2) symmetric limits,
we take the infinite system-size limit
in the Bethe ansatz equations (\ref{finite_BA}) 
under the conditions of $N_1=N_2$ and $N_3=N_4$.
By introducing distribution functions per unit length 
for the continuous rapidities $\alpha$, $\beta$, and $\gamma$,
we obtain the simultaneous integral equations as follows:
\twobeone
\begin{eqnarray}
R_1(\alpha) &=&G_2(\alpha)-\frac{1}{2\pi}\int_{-\infty}^{\infty}G_1(\alpha-\alpha ')R_1(\alpha ')d\alpha '
+\frac{1}{2\pi}\int_{-B}^{B}G_2(\alpha-\beta)R_2(\beta)d\beta, \nonumber \\
R_2(\beta)&=&-\frac{1}{2\pi}\int_{-B}^{B}G_1(\beta-\beta ')R_2(\beta ')d\beta '
+\frac{1}{2\pi}\int_{-\infty}^{\infty}G_2(\beta-\alpha)R_1(\alpha)d\alpha
+\frac{1}{2\pi}\int_{-\infty}^{\infty}G_2(\beta-\gamma)R_3(\gamma)d\gamma, \nonumber\\
R_3(\gamma)&=&-\frac{1}{2\pi}\int_{-\infty}^{\infty}G_1(\gamma -\gamma ')R_3(\gamma ')d\gamma'
+\frac{1}{2\pi}\int_{-B}^{B}G_2(\gamma -\beta)R_2(\beta)d\beta, \label{infinite_BA}
\end{eqnarray}
\onebetwo
where $G_n(x) = n/\pi(1+n^2x^2)$ and $R_1$, $R_2$, and $R_3$ are the
distribution functions per unit length for the rapidities 
$\alpha$, $\beta$, and $\gamma$, respectively. 
The infinite cut-off parameters except for those of $\beta$ 
in the integral equations preserve the relations $N_1=N_2$ and $N_3=N_4$ 
and the cut-off $B$ determines the difference between $N_1$ and $N_3$
and specifies the magnetization $m_z$ through the relation below.
The ground-state energy per site ($\varepsilon_\infty$), 
the momentum $P$ and the magnetization $m_z$ are given by
\begin{eqnarray}
\varepsilon_\infty &=& 1-2\pi\int_{-B}^{B}
G_2(\alpha)R_1(\alpha)d\alpha, \label{infinite_gs}\\
P&=&  N\int_{-B}^{B} K(\alpha)R_1(\alpha)d\alpha, \label{infinite_P}\\ 
m_z&=&2\int_{|\beta|>B}R_2(\beta)d\beta. \label{infinite_mz}
\end{eqnarray}

By solving the coupled integral equations (\ref{infinite_BA}) numerically 
and using the equations (\ref{infinite_gs}) and (\ref{infinite_mz}),
we may again obtain the thermodynamic ground-state energy 
as a function of $m_z$. Since there exist non-systematic errors
in the procedures of discretizations for the integral equations,
it is hard to calculate the  ground-state energies for 
the SU(4) and SU(2) symmetric limits in better accuracy than
$0.1\%$ relative errors to the exact values. 
We have found that, for the calculation of the bulk ground 
state energies, Bethe ansatz method for finite-size systems with
the aid of the extrapolation is more suitable 
than that for the infinite-size system.
Since the SU(4) and SU(2) symmetric points correspond to the limits with
the cut-off parameter $B=\infty$ and 0, respectively, 
we get analytically the asymptotic expressions 
of the thermodynamic ground-state energies ($\varepsilon_\infty$)
around these limits within the lowest-order calculations as follows:
\begin{eqnarray}
\varepsilon_\infty &\sim& 1-\frac{\pi}{4}-\frac{3\log{2}}{2}+ 
\frac{\left(\frac{\pi}{4}\right)^2 m_z}
{\left(\log{\tan{\frac{\pi m_z}{4}}}
\right)^2},\;({\rm for\;}m_z\simeq 0),\\
&\sim& 1-2\log{2}+\log{2}\left(m_z-1\right),\;({\rm for\; }m_z\simeq 1). 
\end{eqnarray}
We also show these functions in Fig.\ref{s_vs_e} by the broken lines.

\subsection{Excitations and incommensurate soft modes}\label{sec_excitation}
In order to obtain excited states by the Bethe ansatz method
for a finite-size system, 
we follow the standard approach to change the distributions of
integer or half-integer quantum numbers 
$J_{\alpha}$, $J_{\beta}$, and $J_{\gamma}$ 
and calculate the elementary excitations with a specified $m_z$.
A removal, from 
the symmetric and continuous distributions of the quantum numbers 
for the ground state, creates a hole and an addition creates a particle.
By creating particle-hole pair excitations corresponding to
the three kinds of quantum numbers $J_\alpha, J_\beta,$ and $J_\gamma$, 
we can discuss three elementary massless excitations in our model.
The calculated excitation spectra as a function of momentum $q$
are shown by the solid lines 
in Fig.\ref{ex48}(0)$\sim$(3) for several values of $m_z$.
We have defined the thermodynamic limit by the extrapolation 
of finite-size data in the same way as for the ground-state energies 
and defined the momentum of the elementary excitations 
by the relative one to the ground state.

Figure \ref{ex48} shows how the excitation spectra change 
between the SU(4) and SU(2) symmetric limits.
In Fig.\ref{ex48}(0), we see that three excitations
have the same velocity $\pi/2$,
which is consistent with the exact result obtained by Sutherland\cite{bethe}.
With the increase of magnetizations, 
the amplitudes and softening wave vectors of the
two small branches of the excitations are decreased continuously
and disappear at the SU(2) limit,
while the largest branch approaches to the SU(2) elementary excitation.
For $m_z=1$, the model reduces to the SU(2) AF Heisenberg model
of the pseudo $\vec{T}$- spins and its elementary excitations have been obtained
to be $\pi \sin q$, which are nothing but the des Cloizeaux and Pearson modes.
It is important to note that for accurate calculations of the
smallest elementary excitations for larger $m_z$, 
we need calculations for larger system sizes
because those excitations correspond to the rearrangements of the 
quantum number $J_\gamma$ and its number $M_3=N(1-m_z)/4$ 
is decreased with the increase of $m_z$.
In other words by using the Fermion model,
this difficulty comes from the fact that
the relative weight of the down spin band to the up spin one is 
decreased with the increase of $m_z$ for a fixed system size.

The wave numbers which correspond to the 
incommensurate soft modes in the elementary excitations 
(Fig.{\ref{ex48}}) are specified by the 
nesting vectors of the Fermi points in Fig.\ref{band} as follows:
$2k_{F,\downarrow}={\pi}\left(1-m_z\right)/2$,
$4k_{F,\downarrow}={\pi}\left(1-m_z\right)$, and
$2k_{F,\uparrow}={\pi}\left(1+m_z\right)/2$.
Since the relation $4k_{F,\downarrow}\equiv 4k_{F,\uparrow}$ (mod. $2\pi$)
is always satisfied, the $4k_{F,\uparrow}$ mode is not independent.
We have found that the excitation spectrum with the softening at 
$q=2k_{F,\downarrow}$ ($2k_{F,\uparrow}$) correspond to
the kink-type excitations of the orbital pseudospin $\vec{T}$ 
in the down (up) spin band and
the spectrum with the softening at $q=4k_{F,\downarrow}$ corresponds to 
the kink-type $\vec{S}$- spin excitations, so-called spinons.

It may be worth to mention that the shift of the momentum of the singularity
in the structure factor is actually observed for the SU(2) spin case.
Dender $et.\,al.$\cite{1/2afexp} have studied the copper
benzoate, Cu(C$_6$D$_5$COO)$_2\cdot$3D$_2$O,
which is known to be the typical spin 1/2 one-dimensional antiferromagnet,
by the neutron scattering experiment and have found
the existence of the field-induced soft modes at the wave vector 
$q=\pi\pm2\pi M(H)/g\mu_B$, which is consistent with the Bethe ansatz
solutions. 

\noindent
\begin{figure}[h] \begin{center}
\leavevmode
\epsfxsize=86mm
\epsffile{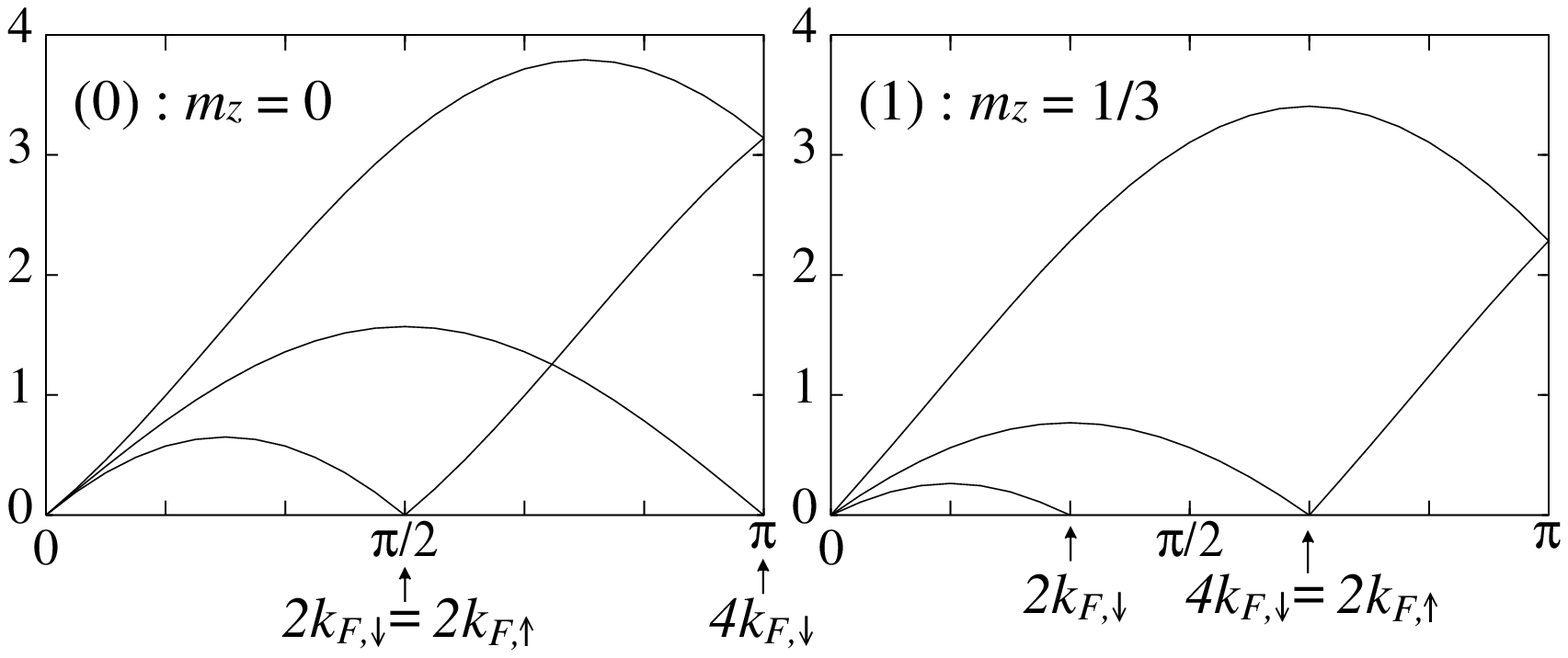}\\
\leavevmode
\epsfxsize=86mm
\epsffile{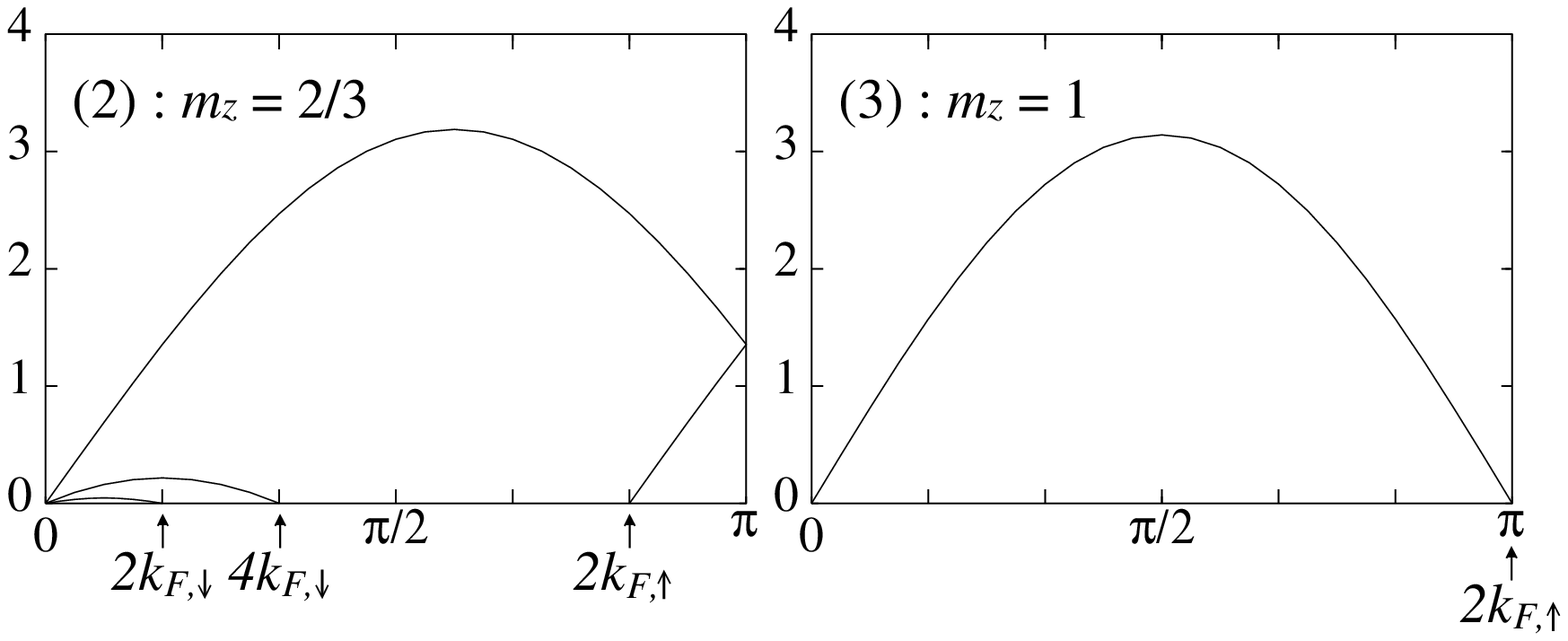}
\end{center} \end{figure}\ \vspace*{-10ex}
\begin{figure}
\caption{ 
Dispersion relations in the infinite-size system limit
for several values of $m_z$.
The vertical and horizontal axis represent the energy and
momentum, respectively.
Below each figure, we label the incommensurate soft modes 
in terms of the original fermion model.
}\label{ex48}
\end{figure}

\section{Crossover phenomena observed in orbital-orbital correlation functions}
\subsection{Correlation functions in real space}
\begin{figure}[htb] \begin{center}
\leavevmode
\epsfxsize=44mm
\epsffile{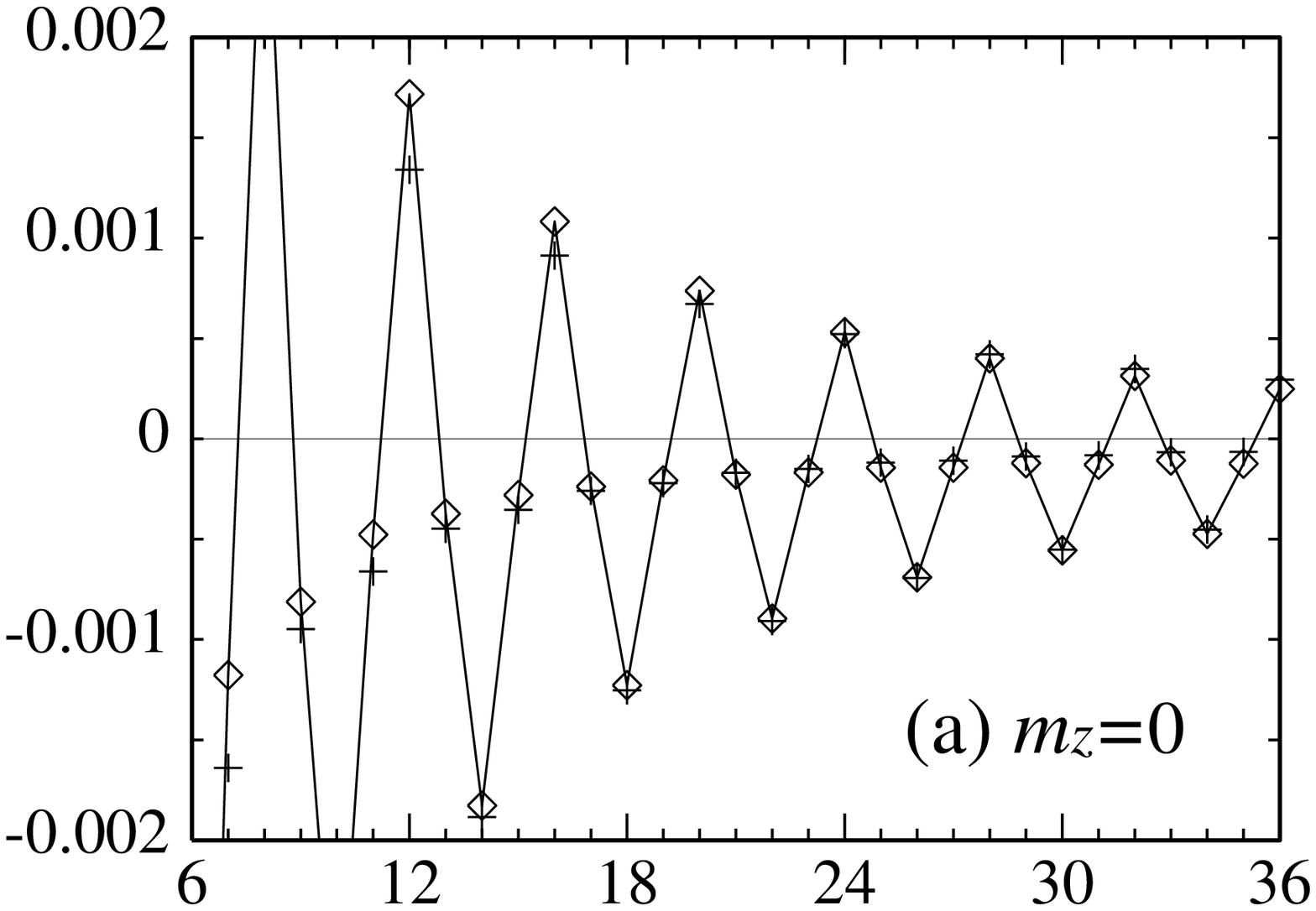}\\
\leavevmode
\epsfxsize=86mm
\epsffile{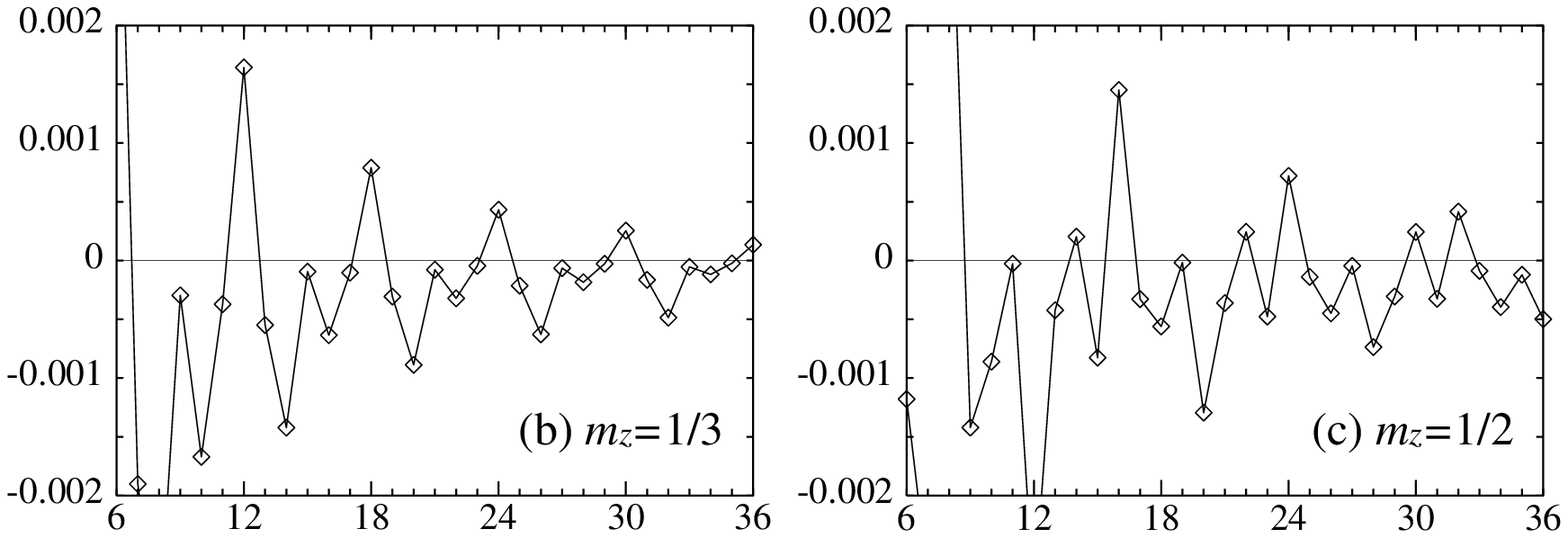}\\
\leavevmode
\epsfxsize=86mm
\epsffile{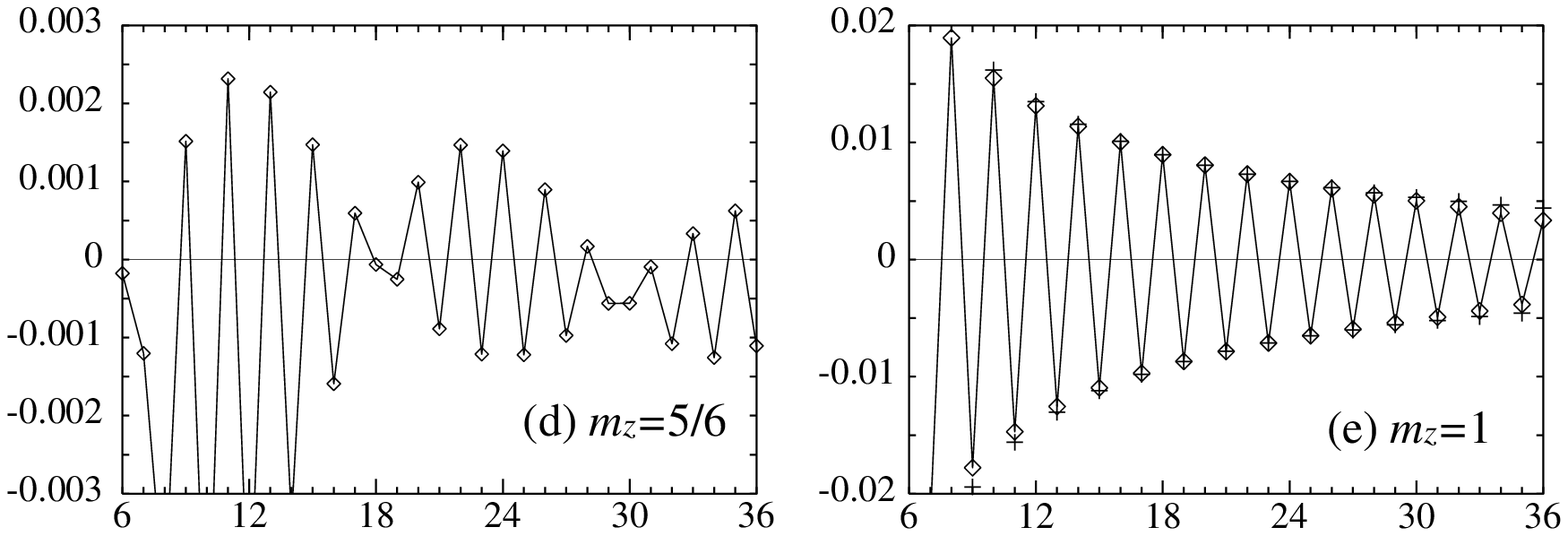}
\caption{Orbital-orbital correlation functions 
$\langle T_i^z T_{i+j}^z\rangle$ as a function of $j$
for the 72-site system with $m_z=0,1/3,1/2,5/6$ and 1.
In figures (a) and (e), we plot the fitting results 
by the symbols $+$.}\label{corr_fig}
\end{center} \end{figure}

In this section, we investigate the orbital-orbital correlation functions
under magnetic fields by the numerical calculations 
using the DMRG method\cite{white}.
Since the DMRG method is more suitable for open boundary conditions (OBC)
than PBC, we employ the OBC in order to carry out the calculations 
for large system sizes with sufficient accuracy.
We consider the 72-site system and obtain the correlation 
functions up to the distance of 36 sites.
As usual we calculate the wave function of the system 
with a specified $S_{tot}^z$
instead of dealing with the model under magnetic fields.
To eliminate the oscillations coming from the open boundaries,
we define the averaged orbital-orbital correlation functions
as follows:
\begin{eqnarray}
\langle T^{z}_{i} T^{z}_{i+j}\rangle \equiv
\frac{1}{4} \sum_{k=0}^{3} \langle {T}^{z}_{i+k} {T}^{z}_{i+k+j} 
\rangle \quad. \label{bulkoocorr}
\end{eqnarray}
We have confirmed that after this averaging procedure the $i$ dependences
of the correlation functions are almost removed.
The obtained correlation functions are shown in Fig.\ref{corr_fig} for
$m_z=0, 1/3, 1/2, 5/6,$ and 1.

For general models with the SU($N$) symmetry, 
the asymptotic behavior of the 
correlation functions have been obtained in a simple form
by Affleck \cite{affleck}.
When we apply his results to our model with 
the SU(4) or SU(2) symmetric limit,
which corresponds to $m_z=0$ or 1, respectively,
the orbital-orbital correlation functions 
behave asymptotically as follows:
\begin{eqnarray*}
\langle {T}^{z}_{i} {T}^{z}_{i+j} \rangle= A_0
\left(
\frac{ \cos{ ({2\pi j}/{N}+\varphi_0)}}{j^{\alpha}}+
\frac{1}{j^{\beta}}\right),
\end{eqnarray*}
where $A_0$ and $\varphi_0$ are non-universal constants
depending on each model, while the critical exponents 
$\alpha=2(1-{1}/{N})$ and $\beta=2$ are universal constants.
For the SU(4) symmetric limit, we try to fit the data from $j=18$ to 32 
with this asymptotic form by the least mean square method.
We do not use the data for $j>32$, because these sites are 
so close to the boundary that they are  affected by the 
oscillations from the open boundary even after the 
averaging procedures. From the fitting we
get the critical exponents to be $\alpha=1.52\pm0.08$, 
and $\beta=2.02\pm0.08$. These exponents are consistent 
with the analytic results by Affleck.
For the other limit with the SU(2) symmetry, 
we similarly estimate the critical
exponents for the oscillating term with wave vector $q=\pi$ 
and non-oscillating term to be $1.04\pm0.01$ and 
$2.5\pm0.6$, respectively, neglecting
the logarithmic corrections. 
These results are again consistent with the exact values 1 and 2.
In this fitting analysis, we have used the data from $j=18$ to 25 
in order to eliminate the boundary effect, 
which is considered to be more noticeable than that
for the SU(4) limit because the critical exponent of
the SU(2) limit is smaller than that of SU(4).

For the intermediate magnetizations between the SU(4) and SU(2) 
symmetric limits, we can not get 
meaningful results concerned with the exponents from a similar fitting 
as the SU(4) or SU(2) limit, since the correlation functions have
more complicated structure as is shown in Fig.\ref{corr_fig}.
However, the correlation functions in real space
show interesting behaviors as the SU(2) symmetric limit is approached.
First of all, let us take a look at Fig.\ref{corr_fig} paying attention
to their periodicities and how they approach to the SU(2) limit. 
For a small value of $m_z$, we see clear periodic 
oscillations with the wave vectors 
$2k_{F,\downarrow}$ and $2k_{F,\uparrow}$ (see Fig.\ref{corr_fig}(a) and (b)), 
which correspond to the incommensurate soft modes of the excitation spectrum
discussed in the previous section.
As we increase magnetic fields, 
the correlation functions approach to the SU(2) one (Fig.\ref{corr_fig}(e))
in such a way as the number of periodic structures observed within 
the 36-site length is decreased and at the same time the SU(2)-like structure 
with the wave vector $\pi$ becomes noticeable (Fig.\ref{corr_fig}(c) and (d)).
These behaviors are consistent with the fact 
in the elementary excitations, that
$2k_{F,\downarrow}$ and $4k_{F,\downarrow}$ soft modes 
vanish and the $2k_{F,\uparrow}$ soft modes continuously
evolve to the des Cloizeaux and Pearson modes at $\pi$ 
with the increase of magnetic fields.
We have also found that for a large value of $m_z$,
the correlation functions are well-illustrated by the 
following pictures where 
the oscillations with the wave vector $\pi$ are modulated by
the slowly-varying 2$k_{F,\downarrow}$.

With the increase of $m_z$, the period of the 2$k_{F,\downarrow}$
oscillations becomes longer and the region, 
where the SU(2)-like behaviors with wave vector $\pi$ are dominant, 
becomes more extended, and in the end for $m_z=1$, the correlation 
functions become the true SU(2) ones.
In this way, the bigger the value of $m_z$ is,
the more distant becomes the asymptotic region, which is governed by the
envelope of 2$k_{F,\downarrow}$ oscillations,
and the more prominent are the SU(2)-like behaviors within the envelope.
Generally speaking, as long as we consider a finite-size system, 
the calculated correlation functions do not reach the 
asymptotic region for a larger value of $m_z$ than some critical 
magnetization, which is an increasing function of the system size,
and the crossover phenomena from the SU(4) to the SU(2) symmetric 
limit take place.
Beyond this characteristic magnetization the SU(2)-like behaviors 
with the slowly-varying envelope described by the $2k_{F,\downarrow}$ 
oscillations are a better picture for the calculated 
correlation functions in a finite length 
than the asymptotic behaviors of the 
$2k_{F,\downarrow}$ and $2k_{F,\uparrow}$ oscillations.

\subsection{Correlation functions in momentum space and asymptotic behaviors}
\begin{figure}[htb]
\begin{center}
\leavevmode
 \epsfxsize=86mm
\epsffile{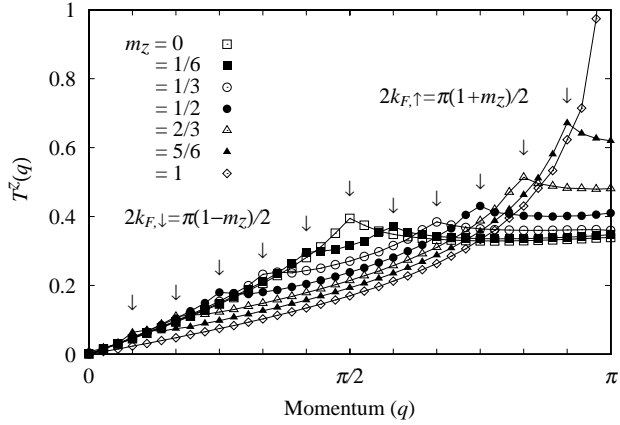}
\caption{ 
Fourier transform of the orbital-orbital correlation functions for
$m_z=0$, 1/6, 1/3, 1/2, 2/3, 5/6, and 1.
We label the wave numbers $2k_{F,\downarrow}$ and $2k_{F,\uparrow}$
by the down arrows ($\downarrow$) for each $m_z$.
$T^z (\pi)$ for $m_z=1$ (which is out of the plot range) is equal to 1.486 .
}\label{tsq}
\end{center}
\end{figure}

Since we want to minutely observe the transition from the SU(4) to 
the SU(2) symmetric limit with the increase of the magnetic field,
next we focus on the asymptotic behavior of the 
correlation functions and concretely estimate the crossover region
between these two limits for a finite-size system with 72-site length.
For this purpose, we discuss the anomalies of the 
structure factors by calculating the fourier transformation 
which is defined by
\begin{equation}
T^{z}(q) \equiv \sum_{j=-N/2+1}^{N/2}\langle T^{z}_{i}
T^{z}_{i+j} \rangle e^{-iqj} \quad .\label{tsq_eq}
\end{equation}
In this definition, we have periodically arranged the 
numerically calculated correlation functions up to 
$N/2$-site length to form the ring with the $N$ sites,
which are less affected by boundary effects,
and defined the structure factors for the ring.
The obtained structure factors for the 72-site system
by changing $m_z$ from 0 to 1 by 1/6 are shown in Fig.\ref{tsq}.
In the figure we see clear two
singular cusp structures at $q=2k_{F,\downarrow}$ 
and $2k_{F,\uparrow}$.
These dominant cusps correspond to the 
field-induced soft modes in the dispersion relations obtained in the 
previous section, and may be interpreted as the orbital excitations 
in the down and up spin bands as illustrated in Fig.\ref{band}.
It is important to note that the $q=4k_{F,\downarrow}$
($\equiv 4k_{F,\uparrow}$ (mod. $2\pi$)) mode does not appear 
in the orbital-orbital correlation functions.
This is because the excitation with $q=4k_{F,\downarrow}$ 
is considered to be an excitation with the spin currents. 
In fact the $4k_{F,\downarrow}$ ($4k_{F,\uparrow}$) momentum transfer 
comes from such excitations as we move two particles from the left
Fermi point of the down (up) spin band to the right one. 
This situation is completely analogous to 
the case of the one-dimensional Hubbard model, where 
the both $2k_{F}$ and $4k_{F}$ oscillating parts appear in 
the density-density correlation function, 
while the $4k_{F}$ part is absent in the spin-spin correlation functions
because the dominant oscillating term with $q=4k_{F}$ 
originates from the holon excitations\cite{kawaholon}. 
According to the conformal field theory\cite{kawacri}, 
the critical exponents $\alpha_1$, $\alpha_2$, and $\alpha_3$
which correspond to the oscillating
terms with $q=2k_{F,\downarrow}$, $4k_{F,\downarrow}$, and
$2k_{F,\uparrow}$, respectively, are obtained as follows:
\begin{eqnarray*}
\alpha_1=\alpha_3 =1+\frac{\xi ^2}{2}\equiv \alpha,\quad 
\alpha_2=2\xi ^2=4(\alpha-1), 
\end{eqnarray*}
where $\xi$ is defined by $\xi (\beta=B)$ and 
$\xi (\beta)$ is obtained by the following
integral equation:
\begin{eqnarray*}
\xi (\beta)=1+\int_{-B}^{B}d\beta '\xi (\beta ')
\Bigg[
\frac{1}{2\pi}\int_{-\infty}^{\infty}dx
\frac{1-e^{-|x|}}{1+e^{|x|}}e^{ix\left(\beta-\beta '\right)}
\Bigg].
\end{eqnarray*}
We calculate these critical exponents as a function of $m_z$,
where $m_z$ is given by Eq.(\ref{infinite_mz}), which only
depends on $B$.
The obtained results are shown in Table \ref{critable}.
In the table, we represent the limit to unity from $m_z <1$ by 1-0. 
The critical exponent $\alpha$ is equal to 1.5 
at $m_z=0$ and 1-0 and continuously changes 
between $0<m_z<1$ with larger values about $1.5\sim 1.6$ 
and jumps to 1 at the SU(2) symmetric point with $m_z=1$.
In Fig.\ref{tsq}, the left and right cusp structures
correspond to the modes with the critical exponent
$\alpha_1$ and $\alpha_3$, respectively.
Though the conformal field theory predicts the same critical
exponent for the both cusps,
the calculated structure factors seem to be quite different,
especially for large $m_z$.
Apparently the left side cusps become less dominant, 
while the right side cusps are more prominent
as the increase of $m_z$.

In order to consider possible reasons, we again employ the 
schematic band structure of the original fermion model 
shown in Fig.\ref{band}.
As the increase of applied magnetic field, it is evident that
the relative weight of the down spin band is diminished as mentioned in
Sec.\ref{sec_excitation}.
It is the reason that the left hand side cusp structures 
are suppressed and their amplitudes become smaller in contrast
to the predictions by the CFT.
By a closer look at the right side cusp structures,
it is found that until the value of $m_z$ less than about
1/3, the cusp shows a little weaker anomaly than that for $m_z=0$, 
which is consistent with the prediction by the CFT, 
while for the region of $m_z>1/2$ the cusps grow rapidly 
and seem to continuously develop into the SU(2) AF 
cusp structure with the critical exponent 1. 

These characteristic features of the fourier transformed 
orbital-orbital correlation functions for the system with 72 sites
are explained by the crossover phenomena due to
the finiteness of the system and do not contradict the CFT predictions.
It is reasonable to consider that in the region $m_z>1/2$ 
the relative weight of the down spin band is so small 
and the 2$k_{F,\downarrow}$ oscillations have such a long period 
that the calculated correlation functions do not reach the asymptotic region
within the 72-site length and show
the properties of the single up spin band
which contribute relatively large weight,
that is the SU(2)-like behavior in the envelope of 
the $2k_{F,\downarrow}$ oscillation.
Therefore the cusp structures seem to continuously evolves into 
the behavior of the SU(2) AF Heisenberg model.
On the other hand for $m_z < 1/3$ both upper and lower split bands have 
significant weight and 2$k_{F,\downarrow}$
and 2$k_{F,\uparrow}$ are not so small that 
the calculated correlation functions are in the asymptotic region 
and they show the asymptotic behaviors of the SU(4) exchange model 
under the magnetic field predicted by the CFT.
This picture is also consistent with the behaviors of the real-space
correlation functions in which the SU(2)-like behaviors become 
noticeable around $m_z=1/2$.

To summarize, for small $m_z$ the orbital-orbital correlation functions
show the oscillations with the wave vectors 2$k_{F,\downarrow}$ and 
2$k_{F,\uparrow}$, 
which give a consistent picture with the prediction by the CFT. 
For the understanding of the region with large $m_z$, however,
we find it useful and appropriate to rewrite 
these two oscillating term as $\cos(2k_{F,\downarrow}j)$
and $\cos(2k_{F,\uparrow}j)=\cos(\pi j)\times \cos(2k_{F,\downarrow}j)$.
The second equation, which comes from the relation 
$2k_{F,\downarrow}+2k_{F,\uparrow}=\pi$
originating from the quarter filling,
gives for the correlation functions of large $m_z$
a simple picture in which 
the SU(2)-like oscillations with the wave vector $\pi$
are modulated by 
the slowly-varying 2$k_{F,\downarrow}$ oscillations.
Following this picture, we may understand straightforwardly
how a finite-size system continuously makes a transition from the SU(4) 
to the SU(2) symmetric limit as we increase magnetic fields.
In particular,
the continuous growth of the cusp structure
into the SU(2) limit shown in Fig.\ref{corr_fig} 
does not contradict the prediction by the
CFT: a jump of the critical exponent
between $m_z=$ 1-0 and 1.

\begin{table}[htb]
\begin{center}
\caption
{The critical exponent $\alpha$ for the magnetizations $m_z$
from 0 to 1 by 1/6.}
\label{critable}
\begin{tabular}{c||c|c|c|c|c|c|c|c} 
$m_z$ & 0   & 1/6  &  1/3 &  1/2 &  2/3 &  5/6 &  1-0 & 1 \\ \hline
$\alpha$    &1.500& 1.562& 1.581& 1.585& 1.573& 1.543& 1.500& 1.000 
\end{tabular}
\end{center}
\end{table}

\section{conclusions and discussions}
In summary, we have investigated the SU(4) exchange model,
which describes the strong quantum fluctuation limit of the
two band Hubbard model, under magnetic fields and 
observed how the system transforms from
the SU(4) symmetric limit with
zero magnetization to the fully polarized SU(2) limit
by using the Bethe ansatz method
and DMRG calculations of the orbital-orbital correlation functions.
First of all, we have kept the trace of the ground-state energy as 
a function of $m_z$ and also obtained the analytic forms
around the SU(4) and SU(2) symmetric limits.
In order to study low energy properties of this model,
we calculated three independent elementary excitations 
by the standard Bethe ansatz technique and found the
field-induced incommensurate soft modes.
Through the DMRG calculations of the orbital-orbital 
correlation functions under magnetic fields,
we have found that the $2k_{F,\downarrow}$ and $2k_{F,\uparrow}$ modes,
which carry the orbital pseudospin currents, 
give rise to dominant cusp structures,
while the 4$k_{F,\downarrow}$ mode, which carries the spin current,
does not appear in the structure factors.
We also conclude that for a finite-length system 
the crossover phenomena take place in the different way predicted by the CFT,
and after that the correlation functions are properly described 
by the SU(2)-like behavior with the slowly-varying envelope of the 
2$k_{F,\downarrow}$ oscillations.
As $m_z$ approaches to unity, the period of envelope becomes longer and longer
and thus the calculated finite-length correlation functions seem to
continuously change into the SU(2) ones for a finite size.
Corresponding to this fact, the cusp structures at the the $2k_{F,\uparrow}$
observed in the structure factors develops into the SU(2) ones continuously
without showing any jumps of the critical exponents.
By comparing the calculated results and the CFT results, we estimate
the crossover region between the SU(4) and the SU(2) symmetric limits
to be $1/3\sim 1/2$ when we consider the relatively short length system
of 72 sites.
It should be noted that from the symmetry of $\vec{S}$- and 
$\vec{T}$- spin degrees of freedom,
we expect similar results for the spin-spin correlation
functions under uniaxial pressures or crystal fields 
produced by single-ion
anisotropies. Under the anisotropies,
one type of orbital of the two is preferred and the four-fold
degenerate bands split into the two doubly degenerate bands
depending on the $T^{z}=\pm 1/2$.

Lastly, we would like to briefly comment on the relevance of the 
apparent difference in the crossover phenomena between the 
calculated correlation functions and
the correlation exponents predicted by the CFT.
Let us take the example of $m_z=5/6$ shown in Fig.\ref{corr_fig}(d).
The $2k_{F,\downarrow}=\pi/12$ oscillation has the period of 24 sites,
and thus it is reasonable to assume that it require a few
hundreds sites to reach the asymptotic region.
From the CFT analysis, we can get the informations about 
the mathematically exact asymptotic behaviors 
for the infinite system size.
In the real materials, however, perfect spin system
cannot be realized and even for very pure systems their
intrinsic length, where the exact one dimensionality 
are kept without any structural defects or impurities, 
would be some hundred atomic lengths at most.
Therefore the results discussed in the present paper
serve a simple example where
the crossover phenomena 
in real systems may be different from the one predicted by the CFT.
In this sense, it is necessary to pay some attention for the application
of the CFT results to realistic finite-length systems.

\acknowledgments
We are grateful to Norio Kawakami for many helpful comments and
discussions.

\end{multicols}
\end{document}